\documentclass[aps,prl,secnumarabic,nobibnotes,twocolumn,superscriptaddress]{revtex4-1}
\usepackage{amsfonts}
\usepackage{mathrsfs}
\usepackage{amsmath}
\usepackage{color}
\usepackage{natbib}
\usepackage{graphicx}
\usepackage{bm}
\usepackage{amssymb}
\usepackage{xspace}
\usepackage{epstopdf}
\usepackage{dcolumn}
\usepackage{multirow}
\usepackage[colorlinks=true, letterpaper=true, pdfstartview=FitV, linkcolor=blue, citecolor=blue, urlcolor=blue]{hyperref}
\usepackage{wrapfig}

\makeatletter

\newcommand{\Rmnum}[1]{\expandafter\@slowromancap\romannumeral #1@}
\makeatother

\begin{document}
\title{Hybrid Nodal Loop Metal: Unconventional Magnetoresponse and Material Realization}

\author{Xiaoming Zhang}
\address{Research Laboratory for Quantum Materials, Singapore University of Technology and Design, Singapore 487372, Singapore}
\address{School of Materials Science and Engineering, Hebei University of Technology, Tianjin 300130, China.}
\author{Zhi-Ming Yu}
\email{zhiming\_yu@sutd.edu.sg}
\address{Research Laboratory for Quantum Materials, Singapore University of Technology and Design, Singapore 487372, Singapore}

\author{Yunhao Lu}
\affiliation{State Key Laboratory of Silicon Materials, School of Materials Science
and Engineering, Zhejiang University, Hangzhou 310027, China}

\author{Xian-Lei Sheng}
\email{xlsheng@buaa.edu.cn}
\address{Research Laboratory for Quantum Materials, Singapore University of Technology and Design, Singapore 487372, Singapore}
\address{Department of Applied Physics, Key Laboratory of Micro-nano Measurement-Manipulation and Physics (Ministry of Education),
Beihang University, Beijing 100191, China}

\author{Hui Ying Yang}
\email{yanghuiying@sutd.edu.sg}
\address{Research Laboratory for Quantum Materials, Singapore University of Technology and Design, Singapore 487372, Singapore}

\author{Shengyuan A. Yang}
\address{Research Laboratory for Quantum Materials, Singapore University of Technology and Design, Singapore 487372, Singapore}

\begin{abstract}
A nodal loop is formed by band crossing along a one-dimensional closed manifold, with each point on the loop a linear nodal point in the transverse dimensions and can be classified as type-I or type-II depending on the band dispersion. Here, we propose a class of nodal loops composed of both type-I and type-II points, which are hence termed as hybrid nodal loops. Based on first-principles calculations, we predict the realization of such loops in the existing electride material  $\rm{Ca_{2}As}$.
For a hybrid loop, the Fermi surface consists of coexisting electron and hole pockets that touch at isolated points for an extended range of Fermi energies, without the need for fine-tuning. This leads to unconventional magnetic responses, including the zero-field magnetic breakdown and the momentum space Klein tunneling observable in the magnetic quantum oscillations, as well as the peculiar anisotropy in the cyclotron resonance.
\end{abstract}
\maketitle

The exploration of new types of quasiparticles in topological band structures has been attracting tremendous attention~\cite{HasanRMP2010,QiRMP2011,ChiuRMP2016,BansilRMP2016}. Due to the reduced symmetry constraints, the kinds of quasiparticles in solids are more abundant than in high-energy physics~\cite{WanPRB2011,XuGPRL2011,YoungPRL2012,WangPRB2012,WangZJ2013,WengHMPRB2016,WengHMPRB20162,Bradlynaaf5037,LvBQ2017bh}. For example, the relativistic Weyl fermions feature an upright Weyl-cone dispersion required by the particle-hole symmetry, which, however, is not a fundamental symmetry in condensed matter and its absence allows the conical dispersion (formed at band-crossings) to be tilted. The degree of the tilt allows the nodal points to be classified into two types ~\cite{XuPRL2015,BernevigNature2015}.
For type-I points, the cone is slightly tilted and the electron-like and hole-like states are occupying different energy ranges; whereas for type-II points, the cone is completely tipped over and the electron-like and hole-like states coexist in energy. Their difference is directly reflected in the distinct Fermi surface topology, which leads to distinct magnetic and transport responses~\cite{NielsenPLB1983,SonPRB2013,YuPRL2016,O'BrienPRL2016,UdagawaPRL2016,TchoumakovPRL2016,KoshinoPRB2016,YanPRL2016,ChangPRL2017,GuanNPJ2017,YangPRL2014}.

The crossing between bands may also form one-dimensional loops in the momentum space~\cite{WengPRB2015,MullenPRL2015,chenNL2015,YuPRL2015,AndrewPRL2015,FangPRB2015,ZhaoJZ2016,XuQNCaP3_2017,ZhangPRB2017}. Each point on the loop can be viewed as a nodal point in the transverse dimensions perpendicular to the loop, hence can be characterized as type-I or type-II. This naturally leads to a classification scheme of nodal loops. Conventional (type-I) loops are composed of type-I points and typically formed by the crossing between an electron-like band and a hole-like band. Li \emph{et al.} proposed the concept of type-II loops~\cite{LiPRB2017}, which consist of only type-II points and are formed by crossing between two electron-like or two hole-like bands. However, there is one remaining possibility: loops composed of \emph{both} type-I and type-II points, which may be termed as hybrid loops.
In previous works, hybrid nodal lines connecting nexus points were predicted in Bernal stacked graphite~\cite{HeikkilNJP2015,HyartPRB2016}, and hybrid loops were noticed in the ScCd-type transition-metal intermetallic materials~\cite{LiPRB2017} as well as some predicted carbon allotropes~\cite{YanarXive2017}. However, unique properties of hybrid loops have not been exposed, and it is urgent to identify an existing material that hosts hybrid loops in order to facilitate the experimental study.

In this work, we present a theory for hybrid nodal loops. We show that the loop typically occurs when one of the crossing bands has a saddle-like dispersion. Based on first-principles calculations, we identify the existing electride material Ca$_2$As as a hybrid nodal-loop metal with three pairs of hybrid loops in the Brillouin zone (BZ) close to the Fermi level. For a hybrid loop, the Fermi surface consists of coexisting electron and hole pockets that touch at isolated points for an extended range of Fermi energies, without the need for fine-tuning. Interestingly, around these touching points, tunneling between the electron and hole pockets would occur in a magnetic field, corresponding to the momentum-space Klein tunneling and forming new types of cyclotron orbits. This in turn leads to unique signatures in magnetic quantum oscillations as well as the peculiar anisotropy in the cyclotron resonance, which can be used to characterize the hybrid nodal-loop phase.

We start by illustrating the essential idea of hybrid loops. Consider a nodal loop formed by the crossing of two bands. The low-energy effective model around a point $P$ on the loop can be expressed as (setting $\hbar=1$)
\begin{equation}\label{Heff}
\mathcal{H}=v_{1}q_{1}\sigma_{x}+v_{2}q_{2}\sigma_{y}+\bm w\cdot \bm q,
\end{equation}
where $\bm q$ is measured from $P$ with $q_{1,2}$ the two components in the plane perpendicular to the loop,
$v_i$'s are the Fermi velocities, and $\sigma_i$'s are the Pauli matrices. The last term in (\ref{Heff}) causes the tilt of the spectrum. Let $\bm w_\bot=(w_1,w_2,0)$ be the component of $\bm w$ in the transverse dimensions, i.e., $q_1$-$q_2$ plane, then the conical dispersion described by (\ref{Heff}) in the $q_1$-$q_2$ plane is (not) tipped over if $|\bm w_\bot|^2$ is greater (less) than $\sqrt{v_1^2 w_1^2+v_2^2 w_2^2}$, and accordingly, the point $P$ is labeled as type-II (type-I).

A hybrid loop emerges when both type-I and type-II points coexist on the loop. Let's consider a simplest model for a single hybrid loop:
\begin{equation}
H=\alpha_x k_x^2+\alpha_y k_y^2 +(\beta_x k_x^2+\beta_y k_y^2 -M)\sigma_x + v_z k_z\sigma_y,
\end{equation}
where $\alpha_i$, $\beta_i$, $M$, and $v_z$ are the model parameters. When $M>0$ (assuming $\beta_{x(y)}>0$), the model describes a nodal loop in the $k_x$-$k_y$ plane, and around each point on the loop, the expanded low-energy model takes the form in (\ref{Heff}). One checks that the loop is type-I (type-II) when the condition $|\alpha_i|/|\beta_i|<1$ ($>1$) holds ($i=x,y$); and a hybrid loop is realized when $(|\alpha_x|-|\beta_x|)(|\alpha_y|-|\beta_y|)<0$.

\begin{figure}
\includegraphics[width=8.8cm]{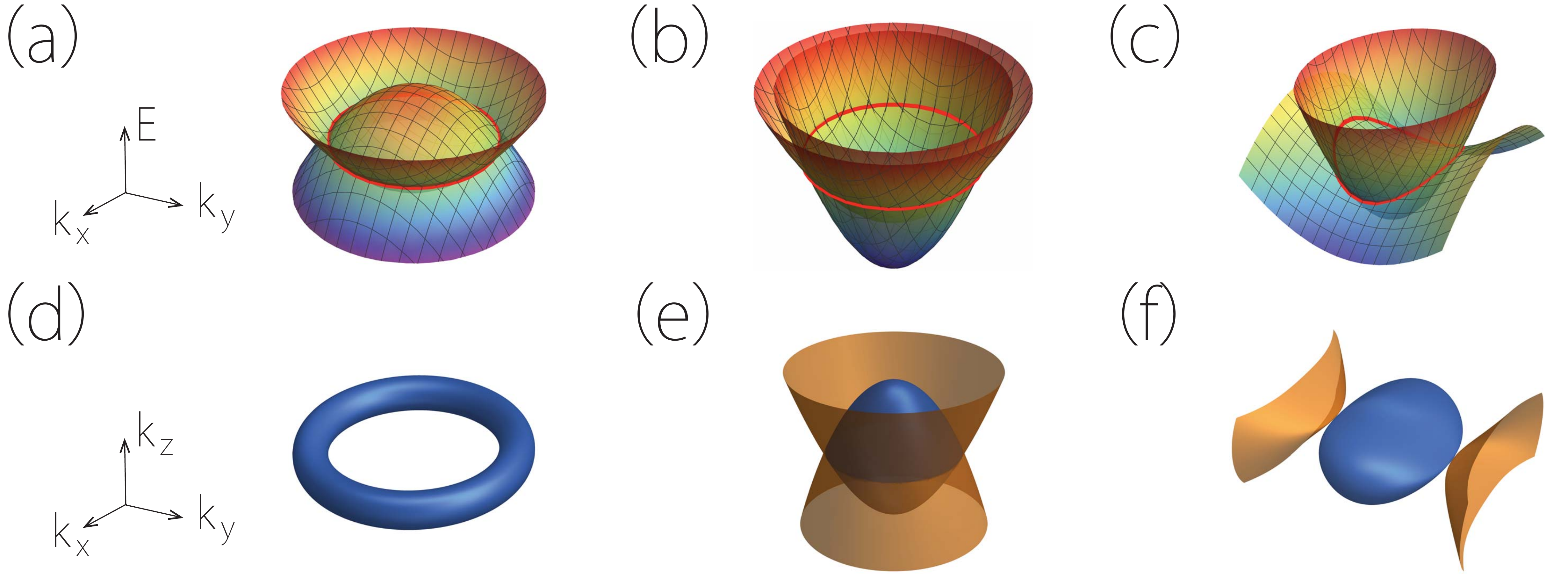}
\caption{Illustration of (a) type-I, (b) type-II, and (c) hybrid nodal loops, with their typical Fermi surfaces shown in (d-f), respectively. Here, the loop is assumed to be in the $k_x$-$k_y$ plane, and blue (orange) color in (d-f) indicates the electron (hole) character of the Fermi surface.
\label{fig1}}
\end{figure}

The three types of loops are illustrated in Fig.~\ref{fig1}. One observes that a hybrid loop would typically occur when one of the crossing bands has a saddle-type dispersion. This leads to two important features. First, a hybrid loop generally has energy variation along the loop, i.e., it spans a range of energy. Second, when the Fermi energy is located in that energy range, the Fermi surface would consist of coexisting electron and hole pockets that are connected by isolated nodal points from the loop. Such touching electron and hole pockets also appear for an isolated type-II nodal point~\cite{XuPRL2015,BernevigNature2015}, which, however, requires the Fermi energy to be exactly tuned to the energy of the nodal point. In comparison, for a hybrid loop, this occurs in a range of energies without the need for fine-tuning. These features give rise to interesting magnetic responses as we shall discuss later.

\begin{figure}
\includegraphics[width=8.8cm]{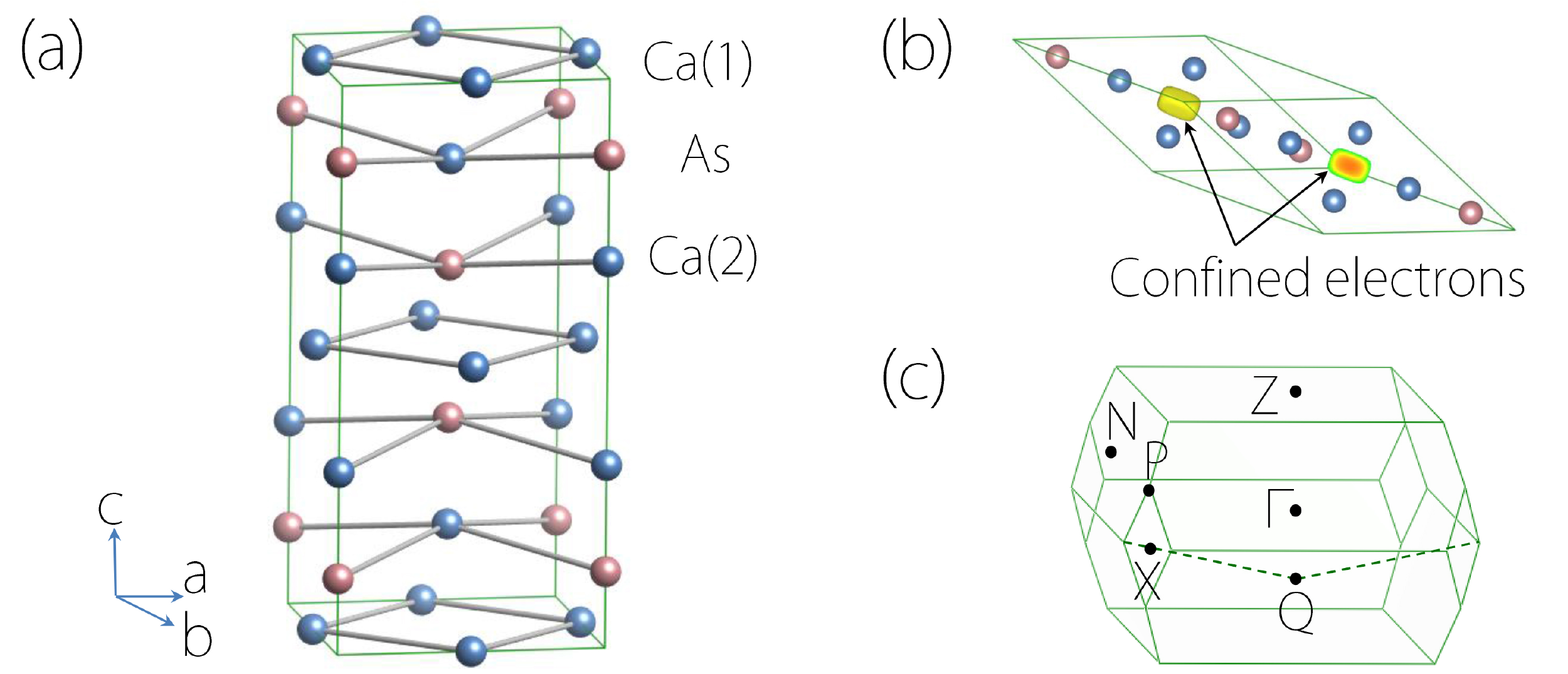}
\caption{Crystal structure of Ca$_2$As in (a) conventianl, and (b) primitive cell representations. (c) The corresponding Brillouin zone. The colored iso-surfaces in (b) (for electron localization function value of $0.85$) indicate the cavities where anionic electrons are confined.
\label{fig2}}
\end{figure}

We identify a concrete material candidate Ca$_2$As that hosts the hybrid loops. Ca$_{2}$As has a body-centered tetragonal structure with the space group $I4/mmm$ (No.~139). Figure~\ref{fig2}(a) shows the conventional cell of Ca$_{2}$As, which contains two units of primitive cell with a hexahedral representation [Fig.~\ref{fig2}(b)]. This material has already been synthesized in experiment~\cite{pearson1985}, and interestingly, it has also been identified as a zero-dimensional electride material~\cite{ZhangPRX2017}, in which there are electrons distributed at interstitial cavities, serving as anions for the structure [see Fig.~\ref{fig2}(b)].

\begin{figure}
\includegraphics[width=8.8cm]{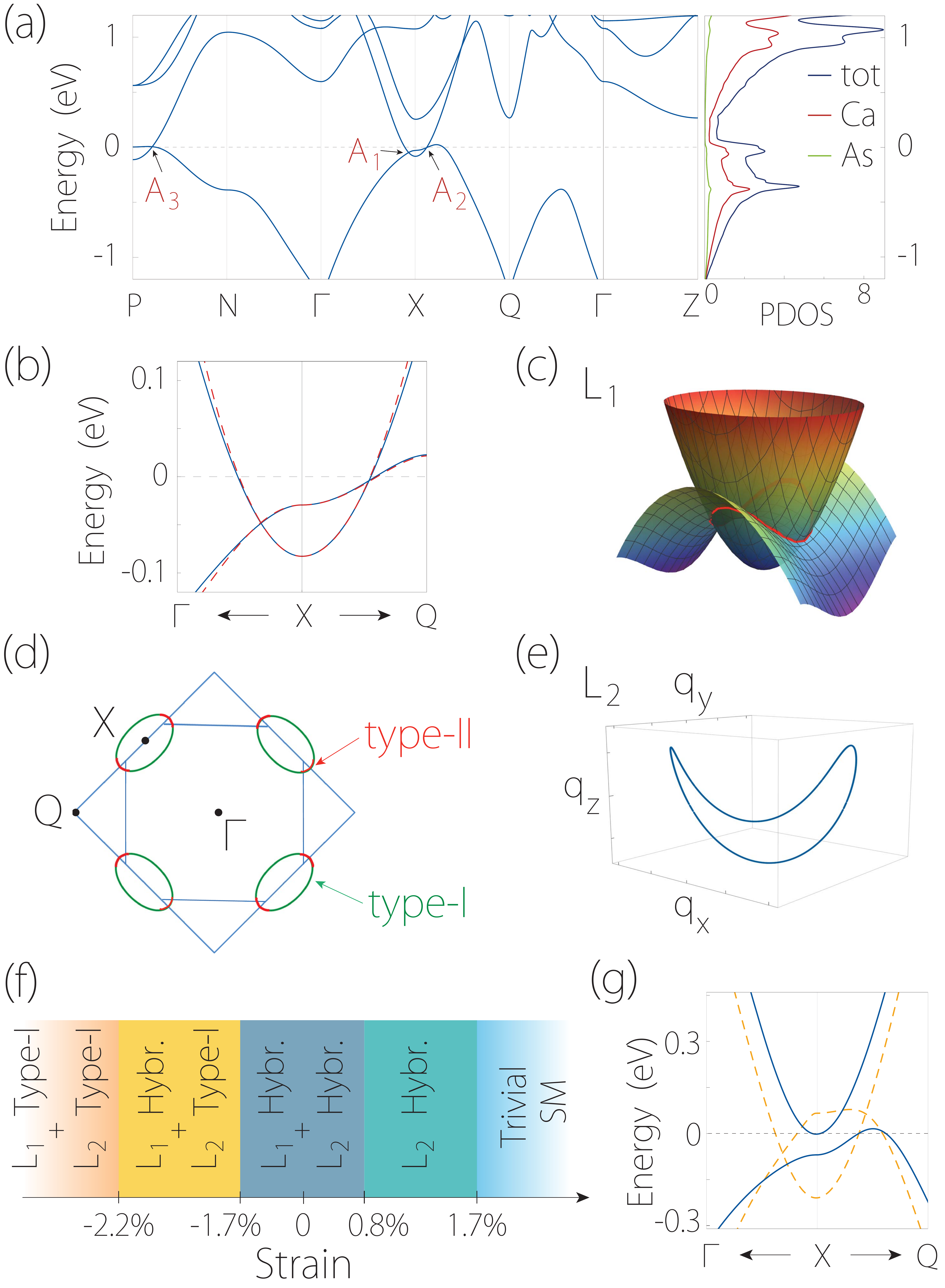}
\caption{(a) Band structure and projected density of states (PDOS) of Ca$_2$As. (b) Enlarged view of band crossing around X (blue line) along with the model fit using Eq.~(\ref{HNR}) (red dashed line). (c) Band dispersion around X in the $k_z = 0$ plane, showing the $L_1$ loop. (d) Location of the two $L_1$ loops in the $k_z=0$ plane. The green (red) color indicates the section with type-I (type-II) points. (e) Shape of the $L_2$ loop around P. (f) Phase diagram of Ca$_2$As under strain. Here, the strain is of biaxial type in the $a$-$b$ plane. (g) Band structure around X under -3\% (dashed line) and 2\% (solid line) strain.
\label{fig3}}
\end{figure}

We performed first-principles calculations on the material properties ~\cite{SuppMater}. The obtained electronic band structure without spin-orbit coupling (SOC) is shown in Fig.~\ref{fig3}(a). It shows a semimetal character with crossing between conduction and valence bands around the X and P points of the BZ. Let's first consider the two crossing points $A_1$ and $A_2$ around X [see Fig.~\ref{fig3}(b)]. First, we find that the two points are not isolated, rather, they belong to the same nodal loop lying in the horizonal ($k_x$-$k_y$) plane. This is explicitly shown in Fig.~\ref{fig3}(c) where we plot the two crossing bands in this plane. In fact, the nodal loop is constrained by the mirror symmetry $\mathcal{M}_z$ to lie in the $k_x$-$k_y$ plane. Second, one notes that the dispersion around $A_1$ and $A_2$ are of distinct types. As observed in the enlarged view in Fig.~\ref{fig3}(b), $A_1$ is type-I whereas $A_2$ is type-II, which indicates that the loop (denoted as $L_1$) is a hybrid loop. This can also be seen in Fig.~\ref{fig3}(c), where one indeed observes that the loop is formed by the crossing between a parabolic band and a saddle-shaped band. In Fig.~\ref{fig3}(d), we show the sections of the loop with type-I or type-II points. And due to the $c_{4z}$ symmetry, there is another symmetry-related hybrid loop located around the ${\rm X}'$ point.

An effective $k\cdot p$ model can be constructed to describe this loop. Without SOC, the two low-energy states at X belong to the $B_{1u}$ and $A_{1g}$ representations of the $D_{2h}$ point group symmetry. Using them as basis, the effective Hamiltonian up to the $k$-quadratic order takes the following form
\begin{eqnarray}
H & = & \left(\begin{array}{cc}
 h_{1} & -iDk_{z}\\
iDk_{z} &  h_{2}
\end{array}\right),\label{HNR}
\end{eqnarray}
where the wave-vector is measured from X, and $h_{1(2)}=m_{1(2)}+a_{1(2)}k_{x}^{2}+b_{1(2)}k_{y}^{2}+c_{1(2)}k_{z}^{2}$. The parameters $m_i$, $a_i$, $b_i$, $c_i$, and $D$ can be fitted from the first-principles result. In Fig.~\ref{fig3}(b), we show the comparison between the first-principles band structure and the model fit. One observes that the model well describes the loop. Interestingly, in the $k_z=0$ plane, the two bands described by $h_1$ and $h_2$ are decoupled, due to their opposite $\mathcal{M}_z$ eigenvalues. For the $h_1$ band, $a_1,b_1>0$ ($a_1=14.3$ eV$\cdot${\rm \AA}$^2$ and $b_1=12.2$ eV$\cdot${\rm \AA}$^2$), so that it is electron-like with the usual parabolic dispersion in the $k_x$-$k_y$ plane. In contrast, for $h_2$, we have $a_2<0$ and $b_2>0$ ($a_2=-7.8$ eV$\cdot${\rm \AA}$^2$ and $b_2=4.4$ eV$\cdot${\rm \AA}$^2$), so its dispersion is of saddle-shape. This analysis unambiguously identifies the loop to be of the hybrid type.

Back to the band structure in Fig.~\ref{fig3}(a), the crossing-point $A_3$ also belongs to another nodal loop (denoted as $L_2$) located around P (and by symmetry, there are four such loops in the BZ).
Due to the reduced symmetry at P, the $L_2$ loop is not confined in a specific plane but has a snake-like shape in $k$-space [see Fig.~\ref{fig3}(e)]. After careful scan of the dispersions, we confirm that the loop is also of hybrid type~\cite{SuppMater}. Thus, Ca$_2$As is indeed a hybrid nodal-loop metal.

Regarding their robustness, when SOC is absent, both $L_1$ and $L_2$ loops are stabilized by the inversion ($\mathcal{P}$) and time reversal ($\mathcal{T}$) symmetries of the system. The presence of $\mathcal{PT}$ symmetry guarantees a quantized Berry phase along any closed 1D path (defining a $\mathbb{Z}_2$ charge). We numerically verifies that for a closed path encircling each loop, its Berry phase is nontrivial ($\pm \pi$), hence the loops are protected against weak perturbations that preserve the symmetry. For the $L_1$ loop, since it is located in the mirror-invariant plane, it enjoys an additional protection by $\mathcal{M}_z$, as the two crossing-bands have opposite $\mathcal{M}_z$-eigenvalues.

Although the loops are robust against weak symmetry-preserving perturbations, a stronger perturbation may remove the loops by pulling the two crossing bands apart, during which a loop shrinks to a point and annihilate. This scenario can be achieved by applying strain. As illustrated in Fig.~\ref{fig3}(f), a rich phase diagram can be realized by simple strain tuning. With biaxial tensile strain in the $a$-$b$ plane, the two loops can be annihilated sequentially. The annihilation of the $L_1$ loop is illustrated in Fig.~\ref{fig3}(g). Under compressive strain, the loops can be turned from the hybrid type to type-I. Thus, strain provides a powerful method to control the nodal loops in the current system.

Including SOC would generally gap the loops. Nevertheless, our calculation shows that SOC-induced gaps in Ca$_2$As for the two loops are less than 3 meV~\cite{SuppMater}, so that the SOC effect can be neglected.

The hybrid nodal loops can be directly probed by mapping out the band structure using ARPES. Like type-I and type-II loops, hybrid loops may also possess drumhead-type surface states~\cite{SuppMater}, which can be detected via scanning tunneling microscopy/spectroscopy. Below, we further discuss the experimental signatures of the hybrid loop in the response under a magnetic field.

As we discussed earlier, for a range of Fermi energies, the Fermi surface for a hybrid loop consists of coexisting electron and hole pockets touching at isolated points that belong to the nodal loop. Such configuration gives rise to pronounced effect of tunneling between the electron and hole pockets when under a magnetic field. The process can be regarded as the momentum-space counterpart of the Klein tunneling effect~\cite{HarrisonPRB2009}. For orbits passing through the touching point, the tunneling probability approaches unity even when the $B$-field strength approaches zero~\cite{O'BrienPRL2016,KoshinoPRB2016}, leading to the zero-field magnetic breakdown which can be observed in magnetic quantum oscillations.

The effect can be understood from a semiclassical picture. The quasiparticle orbits in a $B$-field are quantized according to the Bohr-Sommerfeld condition~\cite{Marder2010}:
\begin{equation}\label{semi}
\ell_B^2 A_{\mathcal C_n^{\pm}}=2\pi\left(n+\nu\right),
\end{equation}
where $\ell_B=\sqrt{1/eB}$ is the magnetic length, $n$ is the integer corresponding to the Landau level (LL) index, and $\nu$  is the phase offset containing the Maslov index and the Berry phase of the orbit. $A_{\mathcal C_n^{\pm}}$ is the area enclosed by the  semiclassical orbit $\mathcal C_n^{\pm}$ in $k$-space, which resides on the intersection between a constant energy surface and a plane perpendicular to the field direction. The $\pm$ in $\mathcal C_n^{\pm}$ indicates that $A$ is a signed area with its sign determined by whether the orbit is in an electron pocket ($\mathcal C^+$) or a hole pocket ($\mathcal C^-$).

The quantized orbits lead to quantum oscillations in the density of states (DOS) periodic in $1/B$, for which the frequency of oscillation is given by the Onsager relation $F_{\pm}=|{\cal{A}}_{\mathcal C_n^{\pm}}|/(2\pi e)$~\cite{onsager1952,Shoenberg1984},
with ${\cal{A}}_{\mathcal C_n^{\pm}}$ denoting the extremal values of $A_{\mathcal C_n^{\pm}}$.
When the electron and hole pockets are well separated in $k$-space, they each contribute its own oscillation frequencies separately. However, when the pockets are close such that their separation is comparable to $\ell_B^{-1}$, there will be sizable tunneling between them, which produces new orbits that are hybridizations of the original electron and hole orbits. As a result, the LLs are reconstructed due to the hybridization between the electron and hole LLs, and new oscillation frequencies corresponding to the hybridized orbits, such as $F_H=|F_+ -F_-|$, would appear.

To explicitly demonstrate the above points, we take the model in Eq.~(\ref{HNR}) for the hybrid loop in Ca$_2$As~\cite{SuppMater}.
In this model, the hybridization of electron and hole orbits can happen when the applied magnetic field is along the $z$- or $x$-direction, but not the $y$-direction. And one notes that the model has mirror symmetry in the $x$-$y$ and $y$-$z$ planes, so the orbit areas must be extremal for $k_z=0$ and for $k_x=0$. Therefore, using the full quantum approach, we numerically solve the LL energy $E_n(B,k_i)$ when $B$ field is along the $i$-direction ($i=z,x$), and compute the partial density of states $\rho(E,B,k_i)=\sum _{n}\delta[E-E_{n}(B,k_i)]$ for $k_i=0$, assuming that it gives the dominant contributions in the oscillation.

\begin{figure}
\includegraphics[width=8.8cm]{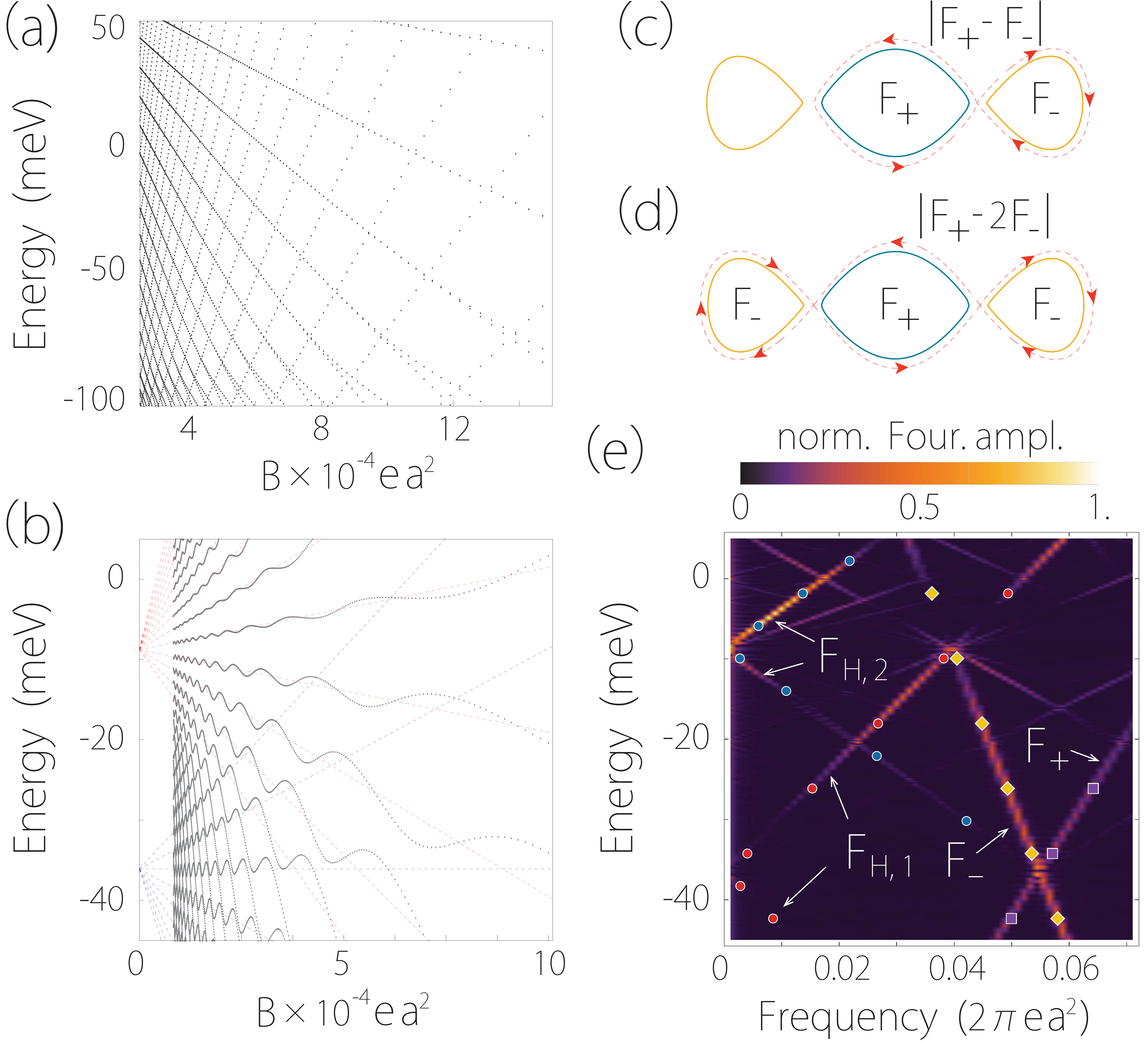}
\caption{(a) Energy spectrum at $k_z=0$ for $B$-field along the $z$ direction. (b) Spectrum at $k_x=0$ for $B$-field along the $x$ direction, where pronounced hybridization (magnetic breakdown) is observed. The dashed lines correspond to the semiclassical
Landau fans for hybridized orbits.
(c,d) Two types of hybridized orbits for $B$-field along the $x$-direction, with characteristic magneto-oscillation frequencies $|F_+-F_-|$ and  $|F_+-2F_-|$ respectively.
(e) Fourier amplitudes of magnetic quantum oscillations for $B$-field along the $x$-direction. Here, the partial density of states $\rho(E, k_x=0)$ are Fourier transformed in the field range $B\in[0.25,~2.5]\times10^{-4}~ e^{-1}a^{-2}$, where $a$ is the length unit.
The data points in the figure are obtained from the Onsager relation. The calculation details are presented in~\cite{SuppMater}
\label{fig4}}
\end{figure}

First, we consider the case for $B$-field along the $z$-direction. Although the electron and hole pockets are touching in the $k_z=0$ plane when the Fermi energy approaches the loop, there is no tunneling between them as they are decoupled in this plane [see Eq.~(\ref{HNR})]. Thus, interestingly, the electron and hole LLs in the $k_z=0$ plane would feature linear crossings versus $B$ regardless of the field strength [see Fig. \ref{fig4}(a)], and such crossings are protected by the mirror symmetry.
Though the tunneling can happen for $k_z\neq 0$, the tunnelling probability wound be suppressed, because when $k_z$ is large, the electron and hole pockets would be well separated in $k$-space.

In sharp contrast, pronounced magnetic breakdown happens for $B$-field along the $x$-direction. Close to the loop energy, the characteristic pattern of the orbits in the $k_x=0$ plane consists of two hole orbits on the two sides separated by an electron orbit in the middle [see Fig.~\ref{fig4}(c)]. Unlike the previous case, here the electron and hole pockets can be coupled by the $B$-field and tunneling between them would occur. We thus expect two additional oscillation frequencies to emerge in the low-frequency range, corresponding to the two new types of hybridized orbits shown in Fig.~\ref{fig4}(c) and \ref{fig4}(d). The first is a hybrid of one electron orbit and one hole orbit, which has a figure-of-eight shape [Fig.~\ref{fig4}(c)]; while the second involves one electron orbit and two hole orbits and has a three-circle shape [Fig.~\ref{fig4}(d)]. Accordingly, the two frequencies are given by $F_{H,1}=|F_+ -F_-|$ and $F_{H,2}=|F_+ -2F_-|$. In Fig.~\ref{fig4}(e), we plot the Fourier amplitude of the oscillation frequencies~\cite{SuppMater}. One clearly observes that for Fermi energy around the hybrid loop, the amplitudes of the original frequencies $F_\pm$ are decreased, and there emerge the $F_{H,1}$ and $F_{H,2}$ frequency peaks in the low-frequency range. Especially, the $F_{H,2}$ peak is most pronounced around $E_F=-9.5$ meV, for which the three orbits are touching, such that the tunneling probability becomes unity.

Experimentally, the quantum oscillation in the DOS manifest in a variety of physical properties, e.g., in the de Haas-van Alphen oscillation in the magnetic susceptibility. The three-circle orbit also leads to the difference in the resonance frequencies in $y$ and $z$ directions by a factor of three. This can be easily understood by noticing that the electron oscillates in the $z$ directions three times as frequently as in $y$. This feature can be detected by comparing the optical absorption peaks for lights polarized in the $y$ and $z$ directions.


\bibliography{HNL_ref}

\begin{thebibliography}{46}%
\makeatletter
\providecommand \@ifxundefined [1]{%
 \@ifx{#1\undefined}
}%
\providecommand \@ifnum [1]{%
 \ifnum #1\expandafter \@firstoftwo
 \else \expandafter \@secondoftwo
 \fi
}%
\providecommand \@ifx [1]{%
 \ifx #1\expandafter \@firstoftwo
 \else \expandafter \@secondoftwo
 \fi
}%
\providecommand \natexlab [1]{#1}%
\providecommand \enquote  [1]{``#1''}%
\providecommand \bibnamefont  [1]{#1}%
\providecommand \bibfnamefont [1]{#1}%
\providecommand \citenamefont [1]{#1}%
\providecommand \href@noop [0]{\@secondoftwo}%
\providecommand \href [0]{\begingroup \@sanitize@url \@href}%
\providecommand \@href[1]{\@@startlink{#1}\@@href}%
\providecommand \@@href[1]{\endgroup#1\@@endlink}%
\providecommand \@sanitize@url [0]{\catcode `\\12\catcode `\$12\catcode
  `\&12\catcode `\#12\catcode `\^12\catcode `\_12\catcode `\%12\relax}%
\providecommand \@@startlink[1]{}%
\providecommand \@@endlink[0]{}%
\providecommand \url  [0]{\begingroup\@sanitize@url \@url }%
\providecommand \@url [1]{\endgroup\@href {#1}{\urlprefix }}%
\providecommand \urlprefix  [0]{URL }%
\providecommand \Eprint [0]{\href }%
\providecommand \doibase [0]{http://dx.doi.org/}%
\providecommand \selectlanguage [0]{\@gobble}%
\providecommand \bibinfo  [0]{\@secondoftwo}%
\providecommand \bibfield  [0]{\@secondoftwo}%
\providecommand \translation [1]{[#1]}%
\providecommand \BibitemOpen [0]{}%
\providecommand \bibitemStop [0]{}%
\providecommand \bibitemNoStop [0]{.\EOS\space}%
\providecommand \EOS [0]{\spacefactor3000\relax}%
\providecommand \BibitemShut  [1]{\csname bibitem#1\endcsname}%
\let\auto@bib@innerbib\@empty
\bibitem [{\citenamefont {Hasan}\ and\ \citenamefont
  {Kane}(2010)}]{HasanRMP2010}%
  \BibitemOpen
  \bibfield  {author} {\bibinfo {author} {\bibfnamefont {M.~Z.}\ \bibnamefont
  {Hasan}}\ and\ \bibinfo {author} {\bibfnamefont {C.~L.}\ \bibnamefont
  {Kane}},\ }\href@noop {} {\bibfield  {journal} {\bibinfo  {journal} {Rev.
  Mod. Phys.}\ }\textbf {\bibinfo {volume} {82}},\ \bibinfo {pages} {3045}
  (\bibinfo {year} {2010})}\BibitemShut {NoStop}%
\bibitem [{\citenamefont {Qi}\ and\ \citenamefont {Zhang}(2011)}]{QiRMP2011}%
  \BibitemOpen
  \bibfield  {author} {\bibinfo {author} {\bibfnamefont {X.-L.}\ \bibnamefont
  {Qi}}\ and\ \bibinfo {author} {\bibfnamefont {S.-C.}\ \bibnamefont {Zhang}},\
  }\href@noop {} {\bibfield  {journal} {\bibinfo  {journal} {Rev. Mod. Phys.}\
  }\textbf {\bibinfo {volume} {83}},\ \bibinfo {pages} {1057} (\bibinfo {year}
  {2011})}\BibitemShut {NoStop}%
\bibitem [{\citenamefont {Chiu}\ \emph {et~al.}(2016)\citenamefont {Chiu},
  \citenamefont {Teo}, \citenamefont {Schnyder},\ and\ \citenamefont
  {Ryu}}]{ChiuRMP2016}%
  \BibitemOpen
  \bibfield  {author} {\bibinfo {author} {\bibfnamefont {C.-K.}\ \bibnamefont
  {Chiu}}, \bibinfo {author} {\bibfnamefont {J.~C.~Y.}\ \bibnamefont {Teo}},
  \bibinfo {author} {\bibfnamefont {A.~P.}\ \bibnamefont {Schnyder}}, \ and\
  \bibinfo {author} {\bibfnamefont {S.}~\bibnamefont {Ryu}},\ }\href@noop {}
  {\bibfield  {journal} {\bibinfo  {journal} {Rev. Mod. Phys.}\ }\textbf
  {\bibinfo {volume} {88}},\ \bibinfo {pages} {035005} (\bibinfo {year}
  {2016})}\BibitemShut {NoStop}%
\bibitem [{\citenamefont {Bansil}\ \emph {et~al.}(2016)\citenamefont {Bansil},
  \citenamefont {Lin},\ and\ \citenamefont {Das}}]{BansilRMP2016}%
  \BibitemOpen
  \bibfield  {author} {\bibinfo {author} {\bibfnamefont {A.}~\bibnamefont
  {Bansil}}, \bibinfo {author} {\bibfnamefont {H.}~\bibnamefont {Lin}}, \ and\
  \bibinfo {author} {\bibfnamefont {T.}~\bibnamefont {Das}},\ }\href@noop {}
  {\bibfield  {journal} {\bibinfo  {journal} {Rev. Mod. Phys.}\ }\textbf
  {\bibinfo {volume} {88}},\ \bibinfo {pages} {021004} (\bibinfo {year}
  {2016})}\BibitemShut {NoStop}%
\bibitem [{\citenamefont {Wan}\ \emph {et~al.}(2011)\citenamefont {Wan},
  \citenamefont {Turner}, \citenamefont {Vishwanath},\ and\ \citenamefont
  {Savrasov}}]{WanPRB2011}%
  \BibitemOpen
  \bibfield  {author} {\bibinfo {author} {\bibfnamefont {X.}~\bibnamefont
  {Wan}}, \bibinfo {author} {\bibfnamefont {A.~M.}\ \bibnamefont {Turner}},
  \bibinfo {author} {\bibfnamefont {A.}~\bibnamefont {Vishwanath}}, \ and\
  \bibinfo {author} {\bibfnamefont {S.~Y.}\ \bibnamefont {Savrasov}},\
  }\href@noop {} {\bibfield  {journal} {\bibinfo  {journal} {Phys. Rev. B}\
  }\textbf {\bibinfo {volume} {83}},\ \bibinfo {pages} {205101} (\bibinfo
  {year} {2011})}\BibitemShut {NoStop}%
\bibitem [{\citenamefont {Xu}\ \emph {et~al.}(2011)\citenamefont {Xu},
  \citenamefont {Weng}, \citenamefont {Wang}, \citenamefont {Dai},\ and\
  \citenamefont {Fang}}]{XuGPRL2011}%
  \BibitemOpen
  \bibfield  {author} {\bibinfo {author} {\bibfnamefont {G.}~\bibnamefont
  {Xu}}, \bibinfo {author} {\bibfnamefont {H.}~\bibnamefont {Weng}}, \bibinfo
  {author} {\bibfnamefont {Z.}~\bibnamefont {Wang}}, \bibinfo {author}
  {\bibfnamefont {X.}~\bibnamefont {Dai}}, \ and\ \bibinfo {author}
  {\bibfnamefont {Z.}~\bibnamefont {Fang}},\ }\href@noop {} {\bibfield
  {journal} {\bibinfo  {journal} {Phys. Rev. Lett.}\ }\textbf {\bibinfo
  {volume} {107}},\ \bibinfo {pages} {186806} (\bibinfo {year}
  {2011})}\BibitemShut {NoStop}%
\bibitem [{\citenamefont {Young}\ \emph {et~al.}(2012)\citenamefont {Young},
  \citenamefont {Zaheer}, \citenamefont {Teo}, \citenamefont {Kane},
  \citenamefont {Mele},\ and\ \citenamefont {Rappe}}]{YoungPRL2012}%
  \BibitemOpen
  \bibfield  {author} {\bibinfo {author} {\bibfnamefont {S.~M.}\ \bibnamefont
  {Young}}, \bibinfo {author} {\bibfnamefont {S.}~\bibnamefont {Zaheer}},
  \bibinfo {author} {\bibfnamefont {J.~C.~Y.}\ \bibnamefont {Teo}}, \bibinfo
  {author} {\bibfnamefont {C.~L.}\ \bibnamefont {Kane}}, \bibinfo {author}
  {\bibfnamefont {E.~J.}\ \bibnamefont {Mele}}, \ and\ \bibinfo {author}
  {\bibfnamefont {A.~M.}\ \bibnamefont {Rappe}},\ }\href@noop {} {\bibfield
  {journal} {\bibinfo  {journal} {Phys. Rev. Lett.}\ }\textbf {\bibinfo
  {volume} {108}},\ \bibinfo {pages} {140405} (\bibinfo {year}
  {2012})}\BibitemShut {NoStop}%
\bibitem [{\citenamefont {Wang}\ \emph {et~al.}(2012)\citenamefont {Wang},
  \citenamefont {Sun}, \citenamefont {Chen}, \citenamefont {Franchini},
  \citenamefont {Xu}, \citenamefont {Weng}, \citenamefont {Dai},\ and\
  \citenamefont {Fang}}]{WangPRB2012}%
  \BibitemOpen
  \bibfield  {author} {\bibinfo {author} {\bibfnamefont {Z.}~\bibnamefont
  {Wang}}, \bibinfo {author} {\bibfnamefont {Y.}~\bibnamefont {Sun}}, \bibinfo
  {author} {\bibfnamefont {X.-Q.}\ \bibnamefont {Chen}}, \bibinfo {author}
  {\bibfnamefont {C.}~\bibnamefont {Franchini}}, \bibinfo {author}
  {\bibfnamefont {G.}~\bibnamefont {Xu}}, \bibinfo {author} {\bibfnamefont
  {H.}~\bibnamefont {Weng}}, \bibinfo {author} {\bibfnamefont {X.}~\bibnamefont
  {Dai}}, \ and\ \bibinfo {author} {\bibfnamefont {Z.}~\bibnamefont {Fang}},\
  }\href@noop {} {\bibfield  {journal} {\bibinfo  {journal} {Phys. Rev. B}\
  }\textbf {\bibinfo {volume} {85}},\ \bibinfo {pages} {195320} (\bibinfo
  {year} {2012})}\BibitemShut {NoStop}%
\bibitem [{\citenamefont {Wang}\ \emph {et~al.}(2013)\citenamefont {Wang},
  \citenamefont {Weng}, \citenamefont {Wu}, \citenamefont {Dai},\ and\
  \citenamefont {Fang}}]{WangZJ2013}%
  \BibitemOpen
  \bibfield  {author} {\bibinfo {author} {\bibfnamefont {Z.}~\bibnamefont
  {Wang}}, \bibinfo {author} {\bibfnamefont {H.}~\bibnamefont {Weng}}, \bibinfo
  {author} {\bibfnamefont {Q.}~\bibnamefont {Wu}}, \bibinfo {author}
  {\bibfnamefont {X.}~\bibnamefont {Dai}}, \ and\ \bibinfo {author}
  {\bibfnamefont {Z.}~\bibnamefont {Fang}},\ }\href@noop {} {\bibfield
  {journal} {\bibinfo  {journal} {Phys. Rev. B}\ }\textbf {\bibinfo {volume}
  {88}},\ \bibinfo {pages} {125427} (\bibinfo {year} {2013})}\BibitemShut
  {NoStop}%
\bibitem [{\citenamefont {Weng}\ \emph
  {et~al.}(2016{\natexlab{a}})\citenamefont {Weng}, \citenamefont {Fang},
  \citenamefont {Fang},\ and\ \citenamefont {Dai}}]{WengHMPRB2016}%
  \BibitemOpen
  \bibfield  {author} {\bibinfo {author} {\bibfnamefont {H.}~\bibnamefont
  {Weng}}, \bibinfo {author} {\bibfnamefont {C.}~\bibnamefont {Fang}}, \bibinfo
  {author} {\bibfnamefont {Z.}~\bibnamefont {Fang}}, \ and\ \bibinfo {author}
  {\bibfnamefont {X.}~\bibnamefont {Dai}},\ }\href@noop {} {\bibfield
  {journal} {\bibinfo  {journal} {Phys. Rev. B}\ }\textbf {\bibinfo {volume}
  {93}},\ \bibinfo {pages} {241202} (\bibinfo {year}
  {2016}{\natexlab{a}})}\BibitemShut {NoStop}%
\bibitem [{\citenamefont {Weng}\ \emph
  {et~al.}(2016{\natexlab{b}})\citenamefont {Weng}, \citenamefont {Fang},
  \citenamefont {Fang},\ and\ \citenamefont {Dai}}]{WengHMPRB20162}%
  \BibitemOpen
  \bibfield  {author} {\bibinfo {author} {\bibfnamefont {H.}~\bibnamefont
  {Weng}}, \bibinfo {author} {\bibfnamefont {C.}~\bibnamefont {Fang}}, \bibinfo
  {author} {\bibfnamefont {Z.}~\bibnamefont {Fang}}, \ and\ \bibinfo {author}
  {\bibfnamefont {X.}~\bibnamefont {Dai}},\ }\href@noop {} {\bibfield
  {journal} {\bibinfo  {journal} {Phys. Rev. B}\ }\textbf {\bibinfo {volume}
  {94}},\ \bibinfo {pages} {165201} (\bibinfo {year}
  {2016}{\natexlab{b}})}\BibitemShut {NoStop}%
\bibitem [{\citenamefont {Bradlyn}\ \emph {et~al.}(2016)\citenamefont
  {Bradlyn}, \citenamefont {Cano}, \citenamefont {Wang}, \citenamefont
  {Vergniory}, \citenamefont {Felser}, \citenamefont {Cava},\ and\
  \citenamefont {Bernevig}}]{Bradlynaaf5037}%
  \BibitemOpen
  \bibfield  {author} {\bibinfo {author} {\bibfnamefont {B.}~\bibnamefont
  {Bradlyn}}, \bibinfo {author} {\bibfnamefont {J.}~\bibnamefont {Cano}},
  \bibinfo {author} {\bibfnamefont {Z.}~\bibnamefont {Wang}}, \bibinfo {author}
  {\bibfnamefont {M.~G.}\ \bibnamefont {Vergniory}}, \bibinfo {author}
  {\bibfnamefont {C.}~\bibnamefont {Felser}}, \bibinfo {author} {\bibfnamefont
  {R.~J.}\ \bibnamefont {Cava}}, \ and\ \bibinfo {author} {\bibfnamefont
  {B.~A.}\ \bibnamefont {Bernevig}},\ }\href@noop {} {\bibfield  {journal}
  {\bibinfo  {journal} {Science}\ }\textbf {\bibinfo {volume} {353}} (\bibinfo
  {year} {2016})}\BibitemShut {NoStop}%
\bibitem [{\citenamefont {Lv}\ \emph {et~al.}(2017)\citenamefont {Lv},
  \citenamefont {Feng}, \citenamefont {Xu}, \citenamefont {Gao}, \citenamefont
  {Ma}, \citenamefont {Kong}, \citenamefont {Richard}, \citenamefont {Huang},
  \citenamefont {Strocov}, \citenamefont {Fang}, \citenamefont {Weng},
  \citenamefont {Shi}, \citenamefont {Qian},\ and\ \citenamefont
  {Ding}}]{LvBQ2017bh}%
  \BibitemOpen
  \bibfield  {author} {\bibinfo {author} {\bibfnamefont {B.~Q.}\ \bibnamefont
  {Lv}}, \bibinfo {author} {\bibfnamefont {Z.~L.}\ \bibnamefont {Feng}},
  \bibinfo {author} {\bibfnamefont {Q.~N.}\ \bibnamefont {Xu}}, \bibinfo
  {author} {\bibfnamefont {X.}~\bibnamefont {Gao}}, \bibinfo {author}
  {\bibfnamefont {J.~Z.}\ \bibnamefont {Ma}}, \bibinfo {author} {\bibfnamefont
  {L.~Y.}\ \bibnamefont {Kong}}, \bibinfo {author} {\bibfnamefont
  {P.}~\bibnamefont {Richard}}, \bibinfo {author} {\bibfnamefont {Y.~B.}\
  \bibnamefont {Huang}}, \bibinfo {author} {\bibfnamefont {V.~N.}\ \bibnamefont
  {Strocov}}, \bibinfo {author} {\bibfnamefont {C.}~\bibnamefont {Fang}},
  \bibinfo {author} {\bibfnamefont {H.~M.}\ \bibnamefont {Weng}}, \bibinfo
  {author} {\bibfnamefont {Y.~G.}\ \bibnamefont {Shi}}, \bibinfo {author}
  {\bibfnamefont {T.}~\bibnamefont {Qian}}, \ and\ \bibinfo {author}
  {\bibfnamefont {H.}~\bibnamefont {Ding}},\ }\href@noop {} {\bibfield
  {journal} {\bibinfo  {journal} {Nature}\ }\textbf {\bibinfo {volume} {546}},\
  \bibinfo {pages} {627} (\bibinfo {year} {2017})}\BibitemShut {NoStop}%
\bibitem [{\citenamefont {Xu}\ \emph {et~al.}(2015)\citenamefont {Xu},
  \citenamefont {Zhang},\ and\ \citenamefont {Zhang}}]{XuPRL2015}%
  \BibitemOpen
  \bibfield  {author} {\bibinfo {author} {\bibfnamefont {Y.}~\bibnamefont
  {Xu}}, \bibinfo {author} {\bibfnamefont {F.}~\bibnamefont {Zhang}}, \ and\
  \bibinfo {author} {\bibfnamefont {C.}~\bibnamefont {Zhang}},\ }\href@noop {}
  {\bibfield  {journal} {\bibinfo  {journal} {Phys. Rev. Lett.}\ }\textbf
  {\bibinfo {volume} {115}},\ \bibinfo {pages} {265304} (\bibinfo {year}
  {2015})}\BibitemShut {NoStop}%
\bibitem [{\citenamefont {Soluyanov}\ \emph {et~al.}(2015)\citenamefont
  {Soluyanov}, \citenamefont {Gresch}, \citenamefont {Wang}, \citenamefont
  {Wu}, \citenamefont {Troyer}, \citenamefont {Dai},\ and\ \citenamefont
  {Bernevig}}]{BernevigNature2015}%
  \BibitemOpen
  \bibfield  {author} {\bibinfo {author} {\bibfnamefont {A.~A.}\ \bibnamefont
  {Soluyanov}}, \bibinfo {author} {\bibfnamefont {D.}~\bibnamefont {Gresch}},
  \bibinfo {author} {\bibfnamefont {Z.}~\bibnamefont {Wang}}, \bibinfo {author}
  {\bibfnamefont {Q.}~\bibnamefont {Wu}}, \bibinfo {author} {\bibfnamefont
  {M.}~\bibnamefont {Troyer}}, \bibinfo {author} {\bibfnamefont
  {X.}~\bibnamefont {Dai}}, \ and\ \bibinfo {author} {\bibfnamefont {B.~A.}\
  \bibnamefont {Bernevig}},\ }\href@noop {} {\bibfield  {journal} {\bibinfo
  {journal} {Nature}\ }\textbf {\bibinfo {volume} {527}},\ \bibinfo {pages}
  {495} (\bibinfo {year} {2015})}\BibitemShut {NoStop}%
\bibitem [{\citenamefont {Nielsen}\ and\ \citenamefont
  {Ninomiya}(1983)}]{NielsenPLB1983}%
  \BibitemOpen
  \bibfield  {author} {\bibinfo {author} {\bibfnamefont {H.~B.}\ \bibnamefont
  {Nielsen}}\ and\ \bibinfo {author} {\bibfnamefont {M.}~\bibnamefont
  {Ninomiya}},\ }\href@noop {} {\bibfield  {journal} {\bibinfo  {journal}
  {Phys. Lett. B}\ }\textbf {\bibinfo {volume} {130}},\ \bibinfo {pages} {389}
  (\bibinfo {year} {1983})}\BibitemShut {NoStop}%
\bibitem [{\citenamefont {Son}\ and\ \citenamefont
  {Spivak}(2013)}]{SonPRB2013}%
  \BibitemOpen
  \bibfield  {author} {\bibinfo {author} {\bibfnamefont {D.~T.}\ \bibnamefont
  {Son}}\ and\ \bibinfo {author} {\bibfnamefont {B.~Z.}\ \bibnamefont
  {Spivak}},\ }\href@noop {} {\bibfield  {journal} {\bibinfo  {journal} {Phys.
  Rev. B}\ }\textbf {\bibinfo {volume} {88}},\ \bibinfo {pages} {104412}
  (\bibinfo {year} {2013})}\BibitemShut {NoStop}%
\bibitem [{\citenamefont {Yu}\ \emph {et~al.}(2016)\citenamefont {Yu},
  \citenamefont {Yao},\ and\ \citenamefont {Yang}}]{YuPRL2016}%
  \BibitemOpen
  \bibfield  {author} {\bibinfo {author} {\bibfnamefont {Z.-M.}\ \bibnamefont
  {Yu}}, \bibinfo {author} {\bibfnamefont {Y.}~\bibnamefont {Yao}}, \ and\
  \bibinfo {author} {\bibfnamefont {S.~A.}\ \bibnamefont {Yang}},\ }\href@noop
  {} {\bibfield  {journal} {\bibinfo  {journal} {Phys. Rev. Lett.}\ }\textbf
  {\bibinfo {volume} {117}},\ \bibinfo {pages} {077202} (\bibinfo {year}
  {2016})}\BibitemShut {NoStop}%
\bibitem [{\citenamefont {O'Brien}\ \emph {et~al.}(2016)\citenamefont
  {O'Brien}, \citenamefont {Diez},\ and\ \citenamefont
  {Beenakker}}]{O'BrienPRL2016}%
  \BibitemOpen
  \bibfield  {author} {\bibinfo {author} {\bibfnamefont {T.~E.}\ \bibnamefont
  {O'Brien}}, \bibinfo {author} {\bibfnamefont {M.}~\bibnamefont {Diez}}, \
  and\ \bibinfo {author} {\bibfnamefont {C.~W.~J.}\ \bibnamefont {Beenakker}},\
  }\href@noop {} {\bibfield  {journal} {\bibinfo  {journal} {Phys. Rev. Lett.}\
  }\textbf {\bibinfo {volume} {116}},\ \bibinfo {pages} {236401} (\bibinfo
  {year} {2016})}\BibitemShut {NoStop}%
\bibitem [{\citenamefont {Udagawa}\ and\ \citenamefont
  {Bergholtz}(2016)}]{UdagawaPRL2016}%
  \BibitemOpen
  \bibfield  {author} {\bibinfo {author} {\bibfnamefont {M.}~\bibnamefont
  {Udagawa}}\ and\ \bibinfo {author} {\bibfnamefont {E.~J.}\ \bibnamefont
  {Bergholtz}},\ }\href@noop {} {\bibfield  {journal} {\bibinfo  {journal}
  {Phys. Rev. Lett.}\ }\textbf {\bibinfo {volume} {117}},\ \bibinfo {pages}
  {086401} (\bibinfo {year} {2016})}\BibitemShut {NoStop}%
\bibitem [{\citenamefont {Tchoumakov}\ \emph {et~al.}(2016)\citenamefont
  {Tchoumakov}, \citenamefont {Civelli},\ and\ \citenamefont
  {Goerbig}}]{TchoumakovPRL2016}%
  \BibitemOpen
  \bibfield  {author} {\bibinfo {author} {\bibfnamefont {S.}~\bibnamefont
  {Tchoumakov}}, \bibinfo {author} {\bibfnamefont {M.}~\bibnamefont {Civelli}},
  \ and\ \bibinfo {author} {\bibfnamefont {M.~O.}\ \bibnamefont {Goerbig}},\
  }\href@noop {} {\bibfield  {journal} {\bibinfo  {journal} {Phys. Rev. Lett.}\
  }\textbf {\bibinfo {volume} {117}},\ \bibinfo {pages} {086402} (\bibinfo
  {year} {2016})}\BibitemShut {NoStop}%
\bibitem [{\citenamefont {Koshino}(2016)}]{KoshinoPRB2016}%
  \BibitemOpen
  \bibfield  {author} {\bibinfo {author} {\bibfnamefont {M.}~\bibnamefont
  {Koshino}},\ }\href@noop {} {\bibfield  {journal} {\bibinfo  {journal} {Phys.
  Rev. B}\ }\textbf {\bibinfo {volume} {94}},\ \bibinfo {pages} {035202}
  (\bibinfo {year} {2016})}\BibitemShut {NoStop}%
\bibitem [{\citenamefont {Yan}\ and\ \citenamefont {Wang}(2016)}]{YanPRL2016}%
  \BibitemOpen
  \bibfield  {author} {\bibinfo {author} {\bibfnamefont {Z.}~\bibnamefont
  {Yan}}\ and\ \bibinfo {author} {\bibfnamefont {Z.}~\bibnamefont {Wang}},\
  }\href@noop {} {\bibfield  {journal} {\bibinfo  {journal} {Phys. Rev. Lett.}\
  }\textbf {\bibinfo {volume} {117}},\ \bibinfo {pages} {087402} (\bibinfo
  {year} {2016})}\BibitemShut {NoStop}%
\bibitem [{\citenamefont {Chang}\ \emph {et~al.}(2017)\citenamefont {Chang},
  \citenamefont {Xu}, \citenamefont {Sanchez}, \citenamefont {Tsai},
  \citenamefont {Huang}, \citenamefont {Chang}, \citenamefont {Hsu},
  \citenamefont {Bian}, \citenamefont {Belopolski}, \citenamefont {Yu},
  \citenamefont {Yang}, \citenamefont {Neupert}, \citenamefont {Jeng},
  \citenamefont {Lin},\ and\ \citenamefont {Hasan}}]{ChangPRL2017}%
  \BibitemOpen
  \bibfield  {author} {\bibinfo {author} {\bibfnamefont {T.-R.}\ \bibnamefont
  {Chang}}, \bibinfo {author} {\bibfnamefont {S.-Y.}\ \bibnamefont {Xu}},
  \bibinfo {author} {\bibfnamefont {D.~S.}\ \bibnamefont {Sanchez}}, \bibinfo
  {author} {\bibfnamefont {W.-F.}\ \bibnamefont {Tsai}}, \bibinfo {author}
  {\bibfnamefont {S.-M.}\ \bibnamefont {Huang}}, \bibinfo {author}
  {\bibfnamefont {G.}~\bibnamefont {Chang}}, \bibinfo {author} {\bibfnamefont
  {C.-H.}\ \bibnamefont {Hsu}}, \bibinfo {author} {\bibfnamefont
  {G.}~\bibnamefont {Bian}}, \bibinfo {author} {\bibfnamefont {I.}~\bibnamefont
  {Belopolski}}, \bibinfo {author} {\bibfnamefont {Z.-M.}\ \bibnamefont {Yu}},
  \bibinfo {author} {\bibfnamefont {S.~A.}\ \bibnamefont {Yang}}, \bibinfo
  {author} {\bibfnamefont {T.}~\bibnamefont {Neupert}}, \bibinfo {author}
  {\bibfnamefont {H.-T.}\ \bibnamefont {Jeng}}, \bibinfo {author}
  {\bibfnamefont {H.}~\bibnamefont {Lin}}, \ and\ \bibinfo {author}
  {\bibfnamefont {M.~Z.}\ \bibnamefont {Hasan}},\ }\href@noop {} {\bibfield
  {journal} {\bibinfo  {journal} {Phys. Rev. Lett.}\ }\textbf {\bibinfo
  {volume} {119}},\ \bibinfo {pages} {026404} (\bibinfo {year}
  {2017})}\BibitemShut {NoStop}%
\bibitem [{\citenamefont {Guan}\ \emph {et~al.}(2017)\citenamefont {Guan},
  \citenamefont {Yu}, \citenamefont {Liu}, \citenamefont {Liu}, \citenamefont
  {Dong}, \citenamefont {Lu}, \citenamefont {Yao},\ and\ \citenamefont
  {Yang}}]{GuanNPJ2017}%
  \BibitemOpen
  \bibfield  {author} {\bibinfo {author} {\bibfnamefont {S.}~\bibnamefont
  {Guan}}, \bibinfo {author} {\bibfnamefont {Z.-M.}\ \bibnamefont {Yu}},
  \bibinfo {author} {\bibfnamefont {Y.}~\bibnamefont {Liu}}, \bibinfo {author}
  {\bibfnamefont {G.-B.}\ \bibnamefont {Liu}}, \bibinfo {author} {\bibfnamefont
  {L.}~\bibnamefont {Dong}}, \bibinfo {author} {\bibfnamefont {Y.}~\bibnamefont
  {Lu}}, \bibinfo {author} {\bibfnamefont {Y.}~\bibnamefont {Yao}}, \ and\
  \bibinfo {author} {\bibfnamefont {S.~A.}\ \bibnamefont {Yang}},\ }\href@noop
  {} {\bibfield  {journal} {\bibinfo  {journal} {npj Quantum Mater.}\ }\textbf
  {\bibinfo {volume} {2}},\ \bibinfo {pages} {23} (\bibinfo {year}
  {2017})}\BibitemShut {NoStop}%
\bibitem [{\citenamefont {Yang}\ \emph {et~al.}(2014)\citenamefont {Yang},
  \citenamefont {Pan},\ and\ \citenamefont {Zhang}}]{YangPRL2014}%
  \BibitemOpen
  \bibfield  {author} {\bibinfo {author} {\bibfnamefont {S.~A.}\ \bibnamefont
  {Yang}}, \bibinfo {author} {\bibfnamefont {H.}~\bibnamefont {Pan}}, \ and\
  \bibinfo {author} {\bibfnamefont {F.}~\bibnamefont {Zhang}},\ }\href@noop {}
  {\bibfield  {journal} {\bibinfo  {journal} {Phys. Rev. Lett.}\ }\textbf
  {\bibinfo {volume} {113}},\ \bibinfo {pages} {046401} (\bibinfo {year}
  {2014})}\BibitemShut {NoStop}%
\bibitem [{\citenamefont {Weng}\ \emph {et~al.}(2015)\citenamefont {Weng},
  \citenamefont {Liang}, \citenamefont {Xu}, \citenamefont {Yu}, \citenamefont
  {Fang}, \citenamefont {Dai},\ and\ \citenamefont {Kawazoe}}]{WengPRB2015}%
  \BibitemOpen
  \bibfield  {author} {\bibinfo {author} {\bibfnamefont {H.}~\bibnamefont
  {Weng}}, \bibinfo {author} {\bibfnamefont {Y.}~\bibnamefont {Liang}},
  \bibinfo {author} {\bibfnamefont {Q.}~\bibnamefont {Xu}}, \bibinfo {author}
  {\bibfnamefont {R.}~\bibnamefont {Yu}}, \bibinfo {author} {\bibfnamefont
  {Z.}~\bibnamefont {Fang}}, \bibinfo {author} {\bibfnamefont {X.}~\bibnamefont
  {Dai}}, \ and\ \bibinfo {author} {\bibfnamefont {Y.}~\bibnamefont
  {Kawazoe}},\ }\href@noop {} {\bibfield  {journal} {\bibinfo  {journal} {Phys.
  Rev. B}\ }\textbf {\bibinfo {volume} {92}},\ \bibinfo {pages} {045108}
  (\bibinfo {year} {2015})}\BibitemShut {NoStop}%
\bibitem [{\citenamefont {Mullen}\ \emph {et~al.}(2015)\citenamefont {Mullen},
  \citenamefont {Uchoa},\ and\ \citenamefont {Glatzhofer}}]{MullenPRL2015}%
  \BibitemOpen
  \bibfield  {author} {\bibinfo {author} {\bibfnamefont {K.}~\bibnamefont
  {Mullen}}, \bibinfo {author} {\bibfnamefont {B.}~\bibnamefont {Uchoa}}, \
  and\ \bibinfo {author} {\bibfnamefont {D.~T.}\ \bibnamefont {Glatzhofer}},\
  }\href@noop {} {\bibfield  {journal} {\bibinfo  {journal} {Phys. Rev. Lett.}\
  }\textbf {\bibinfo {volume} {115}},\ \bibinfo {pages} {026403} (\bibinfo
  {year} {2015})}\BibitemShut {NoStop}%
\bibitem [{\citenamefont {Chen}\ \emph {et~al.}(2015)\citenamefont {Chen},
  \citenamefont {Xie}, \citenamefont {Yang}, \citenamefont {Pan}, \citenamefont
  {Zhang}, \citenamefont {Cohen},\ and\ \citenamefont {Zhang}}]{chenNL2015}%
  \BibitemOpen
  \bibfield  {author} {\bibinfo {author} {\bibfnamefont {Y.}~\bibnamefont
  {Chen}}, \bibinfo {author} {\bibfnamefont {Y.}~\bibnamefont {Xie}}, \bibinfo
  {author} {\bibfnamefont {S.~A.}\ \bibnamefont {Yang}}, \bibinfo {author}
  {\bibfnamefont {H.}~\bibnamefont {Pan}}, \bibinfo {author} {\bibfnamefont
  {F.}~\bibnamefont {Zhang}}, \bibinfo {author} {\bibfnamefont {M.~L.}\
  \bibnamefont {Cohen}}, \ and\ \bibinfo {author} {\bibfnamefont
  {S.}~\bibnamefont {Zhang}},\ }\href@noop {} {\bibfield  {journal} {\bibinfo
  {journal} {Nano letters}\ }\textbf {\bibinfo {volume} {15}},\ \bibinfo
  {pages} {6974} (\bibinfo {year} {2015})}\BibitemShut {NoStop}%
\bibitem [{\citenamefont {Yu}\ \emph {et~al.}(2015)\citenamefont {Yu},
  \citenamefont {Weng}, \citenamefont {Fang}, \citenamefont {Dai},\ and\
  \citenamefont {Hu}}]{YuPRL2015}%
  \BibitemOpen
  \bibfield  {author} {\bibinfo {author} {\bibfnamefont {R.}~\bibnamefont
  {Yu}}, \bibinfo {author} {\bibfnamefont {H.}~\bibnamefont {Weng}}, \bibinfo
  {author} {\bibfnamefont {Z.}~\bibnamefont {Fang}}, \bibinfo {author}
  {\bibfnamefont {X.}~\bibnamefont {Dai}}, \ and\ \bibinfo {author}
  {\bibfnamefont {X.}~\bibnamefont {Hu}},\ }\href@noop {} {\bibfield  {journal}
  {\bibinfo  {journal} {Phys. Rev. Lett.}\ }\textbf {\bibinfo {volume} {115}},\
  \bibinfo {pages} {036807} (\bibinfo {year} {2015})}\BibitemShut {NoStop}%
\bibitem [{\citenamefont {Kim}\ \emph {et~al.}(2015)\citenamefont {Kim},
  \citenamefont {Wieder}, \citenamefont {Kane},\ and\ \citenamefont
  {Rappe}}]{AndrewPRL2015}%
  \BibitemOpen
  \bibfield  {author} {\bibinfo {author} {\bibfnamefont {Y.}~\bibnamefont
  {Kim}}, \bibinfo {author} {\bibfnamefont {B.~J.}\ \bibnamefont {Wieder}},
  \bibinfo {author} {\bibfnamefont {C.~L.}\ \bibnamefont {Kane}}, \ and\
  \bibinfo {author} {\bibfnamefont {A.~M.}\ \bibnamefont {Rappe}},\ }\href@noop
  {} {\bibfield  {journal} {\bibinfo  {journal} {Phys. Rev. Lett.}\ }\textbf
  {\bibinfo {volume} {115}},\ \bibinfo {pages} {036806} (\bibinfo {year}
  {2015})}\BibitemShut {NoStop}%
\bibitem [{\citenamefont {Fang}\ \emph {et~al.}(2015)\citenamefont {Fang},
  \citenamefont {Chen}, \citenamefont {Kee},\ and\ \citenamefont
  {Fu}}]{FangPRB2015}%
  \BibitemOpen
  \bibfield  {author} {\bibinfo {author} {\bibfnamefont {C.}~\bibnamefont
  {Fang}}, \bibinfo {author} {\bibfnamefont {Y.}~\bibnamefont {Chen}}, \bibinfo
  {author} {\bibfnamefont {H.-Y.}\ \bibnamefont {Kee}}, \ and\ \bibinfo
  {author} {\bibfnamefont {L.}~\bibnamefont {Fu}},\ }\href@noop {} {\bibfield
  {journal} {\bibinfo  {journal} {Phys. Rev. B}\ }\textbf {\bibinfo {volume}
  {92}},\ \bibinfo {pages} {081201} (\bibinfo {year} {2015})}\BibitemShut
  {NoStop}%
\bibitem [{\citenamefont {Zhao}\ \emph {et~al.}(2016)\citenamefont {Zhao},
  \citenamefont {Yu}, \citenamefont {Weng},\ and\ \citenamefont
  {Fang}}]{ZhaoJZ2016}%
  \BibitemOpen
  \bibfield  {author} {\bibinfo {author} {\bibfnamefont {J.}~\bibnamefont
  {Zhao}}, \bibinfo {author} {\bibfnamefont {R.}~\bibnamefont {Yu}}, \bibinfo
  {author} {\bibfnamefont {H.}~\bibnamefont {Weng}}, \ and\ \bibinfo {author}
  {\bibfnamefont {Z.}~\bibnamefont {Fang}},\ }\href@noop {} {\bibfield
  {journal} {\bibinfo  {journal} {Phys. Rev. B}\ }\textbf {\bibinfo {volume}
  {94}},\ \bibinfo {pages} {195104} (\bibinfo {year} {2016})}\BibitemShut
  {NoStop}%
\bibitem [{\citenamefont {Xu}\ \emph {et~al.}(2017)\citenamefont {Xu},
  \citenamefont {Yu}, \citenamefont {Fang}, \citenamefont {Dai},\ and\
  \citenamefont {Weng}}]{XuQNCaP3_2017}%
  \BibitemOpen
  \bibfield  {author} {\bibinfo {author} {\bibfnamefont {Q.}~\bibnamefont
  {Xu}}, \bibinfo {author} {\bibfnamefont {R.}~\bibnamefont {Yu}}, \bibinfo
  {author} {\bibfnamefont {Z.}~\bibnamefont {Fang}}, \bibinfo {author}
  {\bibfnamefont {X.}~\bibnamefont {Dai}}, \ and\ \bibinfo {author}
  {\bibfnamefont {H.}~\bibnamefont {Weng}},\ }\href@noop {} {\bibfield
  {journal} {\bibinfo  {journal} {Phys. Rev. B}\ }\textbf {\bibinfo {volume}
  {95}},\ \bibinfo {pages} {045136} (\bibinfo {year} {2017})}\BibitemShut
  {NoStop}%
\bibitem [{\citenamefont {Zhang}\ \emph
  {et~al.}(2017{\natexlab{a}})\citenamefont {Zhang}, \citenamefont {Yu},
  \citenamefont {Sheng}, \citenamefont {Yang},\ and\ \citenamefont
  {Yang}}]{ZhangPRB2017}%
  \BibitemOpen
  \bibfield  {author} {\bibinfo {author} {\bibfnamefont {X.}~\bibnamefont
  {Zhang}}, \bibinfo {author} {\bibfnamefont {Z.-M.}\ \bibnamefont {Yu}},
  \bibinfo {author} {\bibfnamefont {X.-L.}\ \bibnamefont {Sheng}}, \bibinfo
  {author} {\bibfnamefont {H.~Y.}\ \bibnamefont {Yang}}, \ and\ \bibinfo
  {author} {\bibfnamefont {S.~A.}\ \bibnamefont {Yang}},\ }\href@noop {}
  {\bibfield  {journal} {\bibinfo  {journal} {Phys. Rev. B}\ }\textbf {\bibinfo
  {volume} {95}},\ \bibinfo {pages} {235116} (\bibinfo {year}
  {2017}{\natexlab{a}})}\BibitemShut {NoStop}%
\bibitem [{\citenamefont {Li}\ \emph {et~al.}(2017)\citenamefont {Li},
  \citenamefont {Yu}, \citenamefont {Liu}, \citenamefont {Guan}, \citenamefont
  {Wang}, \citenamefont {Zhang}, \citenamefont {Yao},\ and\ \citenamefont
  {Yang}}]{LiPRB2017}%
  \BibitemOpen
  \bibfield  {author} {\bibinfo {author} {\bibfnamefont {S.}~\bibnamefont
  {Li}}, \bibinfo {author} {\bibfnamefont {Z.-M.}\ \bibnamefont {Yu}}, \bibinfo
  {author} {\bibfnamefont {Y.}~\bibnamefont {Liu}}, \bibinfo {author}
  {\bibfnamefont {S.}~\bibnamefont {Guan}}, \bibinfo {author} {\bibfnamefont
  {S.-S.}\ \bibnamefont {Wang}}, \bibinfo {author} {\bibfnamefont
  {X.}~\bibnamefont {Zhang}}, \bibinfo {author} {\bibfnamefont
  {Y.}~\bibnamefont {Yao}}, \ and\ \bibinfo {author} {\bibfnamefont {S.~A.}\
  \bibnamefont {Yang}},\ }\href@noop {} {\bibfield  {journal} {\bibinfo
  {journal} {Phys. Rev. B}\ }\textbf {\bibinfo {volume} {96}},\ \bibinfo
  {pages} {081106} (\bibinfo {year} {2017})}\BibitemShut {NoStop}%
\bibitem [{\citenamefont {Heikkil{\"a}}\ and\ \citenamefont
  {Volovik}(2015)}]{HeikkilNJP2015}%
  \BibitemOpen
  \bibfield  {author} {\bibinfo {author} {\bibfnamefont {T.~T.}\ \bibnamefont
  {Heikkil{\"a}}}\ and\ \bibinfo {author} {\bibfnamefont {G.~E.}\ \bibnamefont
  {Volovik}},\ }\href@noop {} {\bibfield  {journal} {\bibinfo  {journal} {New
  J. of Phys.}\ }\textbf {\bibinfo {volume} {17}},\ \bibinfo {pages} {093019}
  (\bibinfo {year} {2015})}\BibitemShut {NoStop}%
\bibitem [{\citenamefont {Hyart}\ and\ \citenamefont
  {Heikkil\"a}(2016)}]{HyartPRB2016}%
  \BibitemOpen
  \bibfield  {author} {\bibinfo {author} {\bibfnamefont {T.}~\bibnamefont
  {Hyart}}\ and\ \bibinfo {author} {\bibfnamefont {T.~T.}\ \bibnamefont
  {Heikkil\"a}},\ }\href@noop {} {\bibfield  {journal} {\bibinfo  {journal}
  {Phys. Rev. B}\ }\textbf {\bibinfo {volume} {93}},\ \bibinfo {pages} {235147}
  (\bibinfo {year} {2016})}\BibitemShut {NoStop}%
\bibitem [{\citenamefont {Gao}\ \emph {et~al.}()\citenamefont {Gao},
  \citenamefont {Chen}, \citenamefont {Xie}, \citenamefont {Chang},
  \citenamefont {Cohen},\ and\ \citenamefont {Zhang}}]{YanarXive2017}%
  \BibitemOpen
  \bibfield  {author} {\bibinfo {author} {\bibfnamefont {Y.}~\bibnamefont
  {Gao}}, \bibinfo {author} {\bibfnamefont {Y.}~\bibnamefont {Chen}}, \bibinfo
  {author} {\bibfnamefont {Y.}~\bibnamefont {Xie}}, \bibinfo {author}
  {\bibfnamefont {P.}~\bibnamefont {Chang}}, \bibinfo {author} {\bibfnamefont
  {M.~L.}\ \bibnamefont {Cohen}}, \ and\ \bibinfo {author} {\bibfnamefont
  {S.}~\bibnamefont {Zhang}},\ }\href@noop {} {\bibinfo  {journal}
  {arXiv:1707.04576}\ }\BibitemShut {NoStop}%
\bibitem [{\citenamefont {Pearson}(1985)}]{pearson1985}%
  \BibitemOpen
\bibfield  {journal} {  }\bibfield  {author} {\bibinfo {author} {\bibfnamefont
  {W.}~\bibnamefont {Pearson}},\ }\href@noop {} {\bibfield  {journal} {\bibinfo
   {journal} {Z. Kristallogr.}\ }\textbf {\bibinfo {volume} {171}},\ \bibinfo
  {pages} {23} (\bibinfo {year} {1985})}\BibitemShut {NoStop}%
\bibitem [{\citenamefont {Zhang}\ \emph
  {et~al.}(2017{\natexlab{b}})\citenamefont {Zhang}, \citenamefont {Wang},
  \citenamefont {Wang}, \citenamefont {Zhang},\ and\ \citenamefont
  {Ma}}]{ZhangPRX2017}%
  \BibitemOpen
  \bibfield  {author} {\bibinfo {author} {\bibfnamefont {Y.}~\bibnamefont
  {Zhang}}, \bibinfo {author} {\bibfnamefont {H.}~\bibnamefont {Wang}},
  \bibinfo {author} {\bibfnamefont {Y.}~\bibnamefont {Wang}}, \bibinfo {author}
  {\bibfnamefont {L.}~\bibnamefont {Zhang}}, \ and\ \bibinfo {author}
  {\bibfnamefont {Y.}~\bibnamefont {Ma}},\ }\href@noop {} {\bibfield  {journal}
  {\bibinfo  {journal} {Phys. Rev. X}\ }\textbf {\bibinfo {volume} {7}},\
  \bibinfo {pages} {011017} (\bibinfo {year} {2017}{\natexlab{b}})}\BibitemShut
  {NoStop}%
\bibitem [{Sup()}]{SuppMater}%
  \BibitemOpen
  \href@noop {} {\bibinfo  {journal} {See Supplemental Material}\ }\BibitemShut
  {NoStop}%
\bibitem [{\citenamefont {Harrison}\ and\ \citenamefont
  {Sebastian}(2009)}]{HarrisonPRB2009}%
  \BibitemOpen
\bibfield  {journal} {  }\bibfield  {author} {\bibinfo {author} {\bibfnamefont
  {N.}~\bibnamefont {Harrison}}\ and\ \bibinfo {author} {\bibfnamefont {S.~E.}\
  \bibnamefont {Sebastian}},\ }\href@noop {} {\bibfield  {journal} {\bibinfo
  {journal} {Phys. Rev. B}\ }\textbf {\bibinfo {volume} {80}},\ \bibinfo
  {pages} {224512} (\bibinfo {year} {2009})}\BibitemShut {NoStop}%
\bibitem [{\citenamefont {Marder}(2010)}]{Marder2010}%
  \BibitemOpen
  \bibfield  {author} {\bibinfo {author} {\bibfnamefont {M.~P.}\ \bibnamefont
  {Marder}},\ }\href@noop {} {\bibfield  {journal} {\bibinfo  {journal}
  {\emph{Condensed Matter Physics}, 2nd ed.}\ } (\bibinfo {year} {Wiley,
  Hoboken, 2010})}\BibitemShut {NoStop}%
\bibitem [{\citenamefont {Onsager}(1952)}]{onsager1952}%
  \BibitemOpen
  \bibfield  {author} {\bibinfo {author} {\bibfnamefont {L.}~\bibnamefont
  {Onsager}},\ }\href@noop {} {\bibfield  {journal} {\bibinfo  {journal}
  {Philos. Mag.}\ }\textbf {\bibinfo {volume} {43}},\ \bibinfo {pages} {1006}
  (\bibinfo {year} {1952})}\BibitemShut {NoStop}%
\bibitem [{\citenamefont {Shoenberg}(1984)}]{Shoenberg1984}%
  \BibitemOpen
  \bibfield  {author} {\bibinfo {author} {\bibfnamefont {D.}~\bibnamefont
  {Shoenberg}},\ }\href@noop {} {\bibfield  {journal} {\bibinfo  {journal}
  {\emph{Magnetic Oscillations in Metals}}\ } (\bibinfo {year} {Cambridge
  University Press, Cambridge, England, 1984})}\BibitemShut {NoStop}%
\end{thebibliography}%

\end{document}